\documentclass{article}

\newcommand{\beq}{\begin{equation}} 
\newcommand{\eeq}{\end{equation}}
\newcommand{\oma}{\omega_{\rm A}} 
\newcommand{\Om}{\Omega}
\newcommand{\om}{\omega}
\newcommand{\ga}{\;\rlap{\lower 2.5pt
             \hbox{$\sim$}}\raise 1.5pt\hbox{$>$}\;}
\newcommand{\la}{\;\rlap{\lower 2.5pt
             \hbox{$\sim$}}\raise 1.5pt\hbox{$<$}\;}
\newcommand{\aap}[1]{{\em Astron.\ Astrophys.,} {\bf #1}}

\newcommand{\araa}[1]{{\em Ann.\ Rev.\ Astron.\ Astrophys.,} {\bf #1}}
\newcommand{\apj}[1]{{\em Astrophys. J.,} {\bf #1}}

\newcommand{\nat}[1]{{\em Nature,} {\bf #1}}
\newcommand{\sci}[1]{{\em Science,} {\bf #1}}

\newcommand{\pasj}[1]{{\em Publ.\ Astr.\ Soc.\ Japan,} {\bf #1}}

\begin{document}

\centerline{\large\bf Magnetic instability in a differentially rotating star}
\vspace{1\baselineskip}
\centerline{H.C.\ Spruit}
\vspace{1\baselineskip}
\centerline{Max Planck Institute for Astrophysics, Box 1317, 85741 Garching, Germany}
\vspace{3\baselineskip}

{\bf Abstract} 
Instabilities in magnetic fields wound up by differential rotation as reviewed in Spruit (1999) are discussed with some detail and new developments added.  In stellar models which include magnetic torques, the differential rotation tends to accumulate in the gradients in composition. In view of this, instability in a $\mu$-gradient is studied in more detail here, resulting in the detection of a second instability. Its relevance for angular momentum transport is uncertain, however, since it requires high horizontal field gradients and would not operate near the pole. Finally, the possibility is discussed that magnetic instability in  a $\mu$-gradient will lead to {\em layer formation}: the gradients breaking up into small steps of uniform composition and rotation rate. This would enhance the angular momentum transport across inhomogeneous zones, and decrease the rotation rates of the end products of stellar evolution. A recent {\tt astro-ph} submission by Dennisenkov and Pinsonneault proposing a modification of the instability conditions is shown to contain a mathematical error.

\section{Introduction}

\subsection{Strong fields and weak fields}
Magnetic fields are potentially the main mechanism for transport of angular momentum in stars, on account of the high effectiveness of a magnetic field in exerting stresses over large distances, independent of the nature of the fluid in which is embedded (or even across vacuum). 

The `memory-effect' of magnetic fields in HMD, i.e. their dependence on the history of fluid motions, allows large field strengths to build up over time, resulting in torques far larger than can be produced, for example, by the instantaneous overturning action of a hydrodynamic instability. But in principle this same effect also makes magnetic stresses dependent on initial conditions assumed  for the magnetic field of a (differentially rotating) star. In Spruit (2004) I have argued that this dependence in fact leads to a relatively straightforward `bifurcation' of possibilities. 

If the initial field is strong (in some sense that can be quantified), internal magnetic stresses eliminate differential rotation. Such strong fields have their own internal instabilities, causing the field to decay towards a minimum energy state, which depends on the degree to which magnetic helicity is conserved (Braithwaite and Spruit 2004, Braithwaite and Nordlund 2006), resulting in a stable uniformly rotating end state such as  observed in magnetic A-stars. Weak initial fields, on the other hand are wound up by the differential rotation present in the initial state (or developing as a consequence of stellar evolution), producing an azimuthal magnetic field (with respect to the rotation axis) which increases linearly with time until it becomes unstable under it own internal forces. In Spruit (2002, hereafter S02) I have outlined how such instability can feed a field amplification cycle producing a small-scale time dependent `dynamo'. The steady amplitude of this amplification process is then independent of the initial weak field assumed, and depends only on local conditions of differential rotation, thermal and compositional stratification, magnetic and thermal diffusion.

The estimates made in S02 depend critically on the conditions for instability of an azimuthal magnetic field, since it is this instability that closes the field-amplification loop. Fortunately, at least the linear stability conditions for an azimuthal magnetic field can be analyzed in great detail since the instabilities are intrinsically local in a meridional plane (though global in the azimuthal direction). Instability conditions were derived in an appendix in Spruit (1999, hereafter S99), on the basis of a general dispersion relation derived by Acheson (1978). They include the effects of stabilizing thermal stratification and gradients in composition, thermal and magnetic diffusion. 

\subsection{Observational tests}

In a star like the Sun the composition is still nearly homogeneous, with exception of the Hydrogen burning core and the effect of gravitational settling of Helium below the convection zone. In this case, the stabilizing effect of the thermal stratification, though strongly relaxed by thermal diffusion, is the dominant influence on instability. The slow and nearly uniform rotation of the radiative interior of the Sun is compatible with angular momentum transport by magnetic fields according to the dynamo estimate of S02. 

The internal rotation of the Sun is a key test case for any theory of angular momentum transport in stars, first of all because it rotates so slowly that most hydrodynamic transport mechanism fail to explain the observed uniformity of its internal rotation. This led to the early suggestion (Spruit et al. 1983) that magnetic fields are the dominant transport mechanism in the radiative interior of the Sun.  A possible exception is transport by internal gravity waves excited by the convection zone (Charbonnel and Talon 2005). Secondly, the Sun is also the only star for which a value for the spindown torque, independent of the rotation rate, exists (from measurements in the solar wind). Some prescriptions for internal torques commonly used in stellar evolution codes (`with rotation') fail to reproduce the internal rotation of the Sun by large factors. 

Observations of rotation rates in various stages of stellar evolution give a confusing picture, however, and it seems unlikely that a single mechanism can fit all. Initial spins of pulsars around 10 ms are consistent with the magnetic torques of S02 (Heger et al. 2005). The collapsar model for gamma-ray bursts, on the other hand, requires much larger amounts of angular momentum to survive in a pre-supernova core. Thus any single, deterministic mechanism for angular momentum transport can not fit both pulsar rotation and GRB. 

The rotation of white dwarfs can be explored as an indication of rotation rates in the cores of giants, and hence indirectly also of internal angular momentum transport during the giant phase. The observed rotation of WD scatters widely, however. In Spruit (1998) I have argued that the rotation of WD is in fact not a remnant of the initial angular momentum of the star, but a result of small random asymmetries in the mass loss on the AGB. A similar effect (`birth kicks') has been proposed for the rotation of neutron stars (Spruit and Phinney 1998). If it contributes significantly to pulsar rotation, the contribution from the initial angular momentum must be less, implying that the prescription in S02 in fact still underestimates the internal torques in a star. The presence of a kick contribution to pulsar rotation can actually be tested observationally by correlation of spin axes and proper motions of pulsars (Spruit and Phinney 1998, Wang et al. 2006). 

The rotation rates of horizontal branch stars are a potential source of information on angular momentum transport in stars on the giant branch (de Medeiros 2004 and references therein). The observation that the more compact blue HB stars tend to rotate more slowly than red HB stars, however, is a challenge for any mechanism for the angular momentum transport in their giant progenitors.

Inferences about internal torques in stars have been made from patterns of rotation in young clusters of different ages. Differential rotation between core and envelope surviving for $10^{7-8}$yr fits these data (Dennisenkov and Pinsonneault 2006 and references therein). The strong internal  coupling implied by the dynamo estimate of S02, on the other hand, is not compatible with this interpretation. Dennisenkov and Pinsonneault (2006) propose a modification of the stability conditions in S99, resulting in lower magnetic torques. The mathematics of this modification contains an error, however. This is explained below in a paragraph following eq.\ (\ref{dr3}).

Calculations of the internal rotation of evolving stars including the magnetic torque prescriptions in S02 have been done by Heger et al.\ (2005), Maeder and Meynet (2005), and Yoon and Langer (2006). They show that regions of homogeneous composition quickly become uniformly rotating, with gradients in rotation rate accumulating in the composition gradients. Angular momentum transport across  $\mu$-gradients is thus the critical factor for the internal rotation of evolved stars. For this reason the instabilities in a composition gradient are scrutinized a bit closer here than was done in the appendix of S99. This is done in section \ref{compo}.

\section{Instabilities without a $\mu$-gradient}
\label{nog}
In this section the mathematics leading to the stability conditions in S99 is spelled out in a bit more detail, and expanded to include a discussion of the case when the magnetic frequency is larger than the rotation rate.

Near the pole, and in the limit $v_{\rm A}\ll c_{\rm s}$ Acheson's dispersion reduces to eq. (A8) in S99, and can be written as

\begin{eqnarray} 
&\oma^2(\Om m+\omega)+{m^2\over 2}\left[\omega-{\oma^2\over \omega+i\eta s^2}-{l^2\over n^2}{N^2\over\omega+i\kappa s^2/\gamma}\right] [\omega(\omega+i\eta s^2)-\oma^2] - \nonumber  \\
& \left[2\Om m+{\oma^2\over \omega+i\eta s^2}(p+1)\right] [\omega m(\omega+i\eta s^2)+\oma^2]=0.\label{dr1}
\end{eqnarray}

Here $\omega$ is the (complex) frequency of the perturbation, $l,n$ the horizontal and vertical components of the wave number, $m$ the azimuthal order, $s^2=l^2+n^2$, $p$ the index of the variation of the field strength with distance $\varpi$ from the axis $p={\rm d}\ln B/{\rm d}\ln\varpi$, and $\oma$ the characteristic magnetic frequency $\oma=v_{\rm A}/\varpi$. $N$ is the buoyancy frequency, $\eta$ the magnetic diffusivity, $\kappa$ the thermal diffusivity. Viscosity has been neglected.

In all of the following, I assume $m>0$. Axisymmetric perturbations have to be analyzed separately from (\ref{dr1}), but they are unlikely to be very interesting.

Equation (\ref {dr1}) is of fifth order in $\omega$, hence analytically tractable only in limiting cases. Fortunately, the parameter space most likely to be relevant in stars has an ordering of time scales: 
\beq N\gg\Om\gg\oma.\label{order} \eeq 
[The latter of these inequalities is not immediately obvious; this is discussed separately in section \ref{lowom} below.] 
To see what the limit $\Om\gg\oma$ implies, introduce the following dimensionless quantities:
\beq 
\alpha=\omega{\Om\over \oma^2},\qquad h={s^2\eta\Om\over\oma^2};\qquad k={s^2\kappa\Om\over\gamma\oma^2},
\eeq
\beq
\qquad \epsilon={\oma\over\Om}, \qquad Q={l^2\over n^2}{N^2\over\oma^2},
\eeq
(note difference of definition of $h,k$ compared with S99).
After multiplication by $(\alpha+ih)(\alpha+ik)$ eq.\ \ref{dr1} then reads (without assuming $\epsilon$ to be small yet):
\begin{eqnarray} 
&(m+\alpha\epsilon^2)(p-1)(\alpha+ih)(\alpha+ik)  \nonumber  \\
&+{m^2\over 2} [\alpha\epsilon^2 (\alpha+ih)(\alpha+ik)-(\alpha+ik) -Q(\alpha+ih)]  [\alpha\epsilon^2(\alpha+ih)-1] \nonumber  \\
&- [2m(\alpha+ih)+p+1] [m(\alpha+ih)+1] (\alpha+ik)=0. \label{dr2}
\end{eqnarray}
All physical effects (buoyancy, magnetic energy, thermal and magnetic diffusion),  contribute to the solution of this equation if all terms and factors, including the frequency $\alpha$, are of the same order. One of the things this implies is that $Q$ must be $O(1)$, or
\beq l/n\sim N/\oma\ll 1,\eeq
that is to say, the radial scale of the mode must be much smaller than its horizontal scale.
If $l/n$ is not small, the term involving $Q$ dominates, and must be balanced in the first square bracket in (\ref{dr2}) by a suitably large $\omega$. The solutions for such wavenumbers are thus buoyancy waves (gravity waves), $\omega^2=(l^2/s^2)N^2$. For somewhat smaller values of $Q$, but still $\gg 1$, a balance is possible with frequencies around the rotation rate $\Om$; these rotationally modified are buoyancy waves are called inertial-gravity waves. With the ordering of time scales (\ref{order}) they are unaffected by the magnetic field. To find the modes in which the magnetic field plays a role, one must have $Q\sim O(1)$. This is the formal justification of the heuristic derivation of the minimum vertical wavenumber in S99 (eq.\ 44). Note that it does not involve the rotation rate $\Om$. 

Magnetic and thermal diffusion become important for wavenumbers such that $h,k\sim O(1)$, respectively, or
\beq 
n^2\sim {\oma^2\over\eta\Om}; \qquad n^2\sim {\oma^2\over\kappa\Om}.\label{wn}
\eeq
The typical value of $\alpha$ for a solution of (\ref{dr2}) is of order unity, and thus the growth rate $\sigma=-{\rm Im}(\omega)$ of an instability of order $\sim\oma^2/\Om$. Hence we can interpret (\ref{wn}) as stating that the effect of magnetic diffusion on instability becomes important when the growth time is of the same order as the diffusive time scale $1/(n^2\eta)$ (and similar for thermal diffusion $\kappa$). The rotation rate therefore {\em does} have an effect on the high end of the wavenumber range, where diffusion becomes important. This agrees with the heuristic argument in S02. 

Now seeking solutions for the case $\epsilon \ll 1$, but keeping all other quantities in (\ref{dr2}) of order unity, we can expand (\ref{dr2}) in powers of $\epsilon^2$. It appears only in the combination $\alpha\epsilon^2$, with the result that at the lowest order in $\epsilon$, (\ref{dr2}) is only of third order in $\alpha$ (eq.\ A18 in S99):
\begin{eqnarray}
&m(p-1)(\alpha+ih)(\alpha+ik)+{m^2\over 2}[\alpha(1+Q)+ih+ik]  \nonumber  \\
&- [2m(\alpha+ih)+p+1] [m((\alpha+ih)+1] (\alpha+ik)=0.\label{dr3}
\end{eqnarray}
The two remaining roots of the original fifth order equation have disappeared because they do not scale as $\omega\sim\oma^2/\Om$. 

The procedure to derive instability conditions from this is now as follows. Since the coefficients in (\ref{dr3}) are complex, the solutions $\alpha$ are complex as well. Roots with ${\rm Im}(\alpha)>0$ decay as $\exp[-{\rm Im}(\alpha)]\exp[i{\rm Re}(\alpha)]$, hence correspond to stability. Roots with ${\rm Im}(\alpha)<0$ correspond to instability. The boundary between stable and unstable modes is at ${\rm Im}(\alpha)=0$. To find the instability condition, we thus set ${\rm Im}(\alpha)=0$ and take Re and Im parts of (\ref{dr3}), yielding two equations for the real quantity $\alpha$ that have to be satisfied {\em simultaneously}. Together, they determine not only the value of $\alpha$, but also a relation between the parameters that has to be satisfied; this relation defines the boundary between stable and unstable roots.

In their discussion of the dispersion relation (\ref{dr2}) Dennisenkov and Pinsonneault (2006) misinterpret this procedure. By assuming $\alpha$ to be real (as above) and eliminating the highest power of $\alpha$ between the real and imaginary parts of (\ref{dr2}) they obtain an equation of lower order. They proceed to obtain {\em complex} solutions of this equation, apparently motivated by the fact that complex values of the mode frequency indicate instability. This is incorrect since the separation into real and imaginary parts used holds only for Im($\alpha=0$), and it ignores the fact that there are two equations to be satisfied.

The imaginary part of (\ref{dr3}) can be written as:
\beq 
-2m^2\alpha^2-4m\alpha-4{h\over k}(m^2\alpha^2+m\alpha)+{m^2\over 2}(Q{h\over k}+1) - (p+1) + 2m^2h^2=0. \label{dr3i}
\eeq
It is not necessary to consider the general case where both diffusivities are similar. In a stellar interior the thermal diffusivity is generally much larger than the magnetic diffusivity. In taking the limit $h/k=\gamma\eta/\kappa\ll 1$ in (\ref{dr3i}), the term involving $Q$ must be kept, however. It represents the stabilizing effect of the stratification, proportional to the square of the aspect ratio $l/s$ of the perturbations. The condition $Q\eta/\kappa\sim O(1)$ determines the range of aspect ratios that is relevant for the effect of the stable stratification on the instability. Due to the factor $\eta/\kappa$ multiplying $Q$ compared with the adiabatic case $\eta=\kappa=0$, we can already conclude that the range of unstable aspect ratios is now {\em larger} than in the adiabatic case. Effectively,  thermal diffusion reduces the buoyancy $N^2$ by a factor  $\eta/\kappa$. This is very similar to the effect it has on purely hydrodynamic instabilities (Zahn 1974). Thus (\ref{dr3i}) becomes:
\beq 
-2m^2\alpha^2-4m\alpha+{m^2\over 2}(Q{h\over k}+1) - (p+1) + 2m^2h^2=0. \qquad (\eta/\kappa\ll 1) \label{dr3i1}
\eeq
This can be written as
\beq
2-2(m\alpha+1)^2+{m^2\over 2}-(p+1)+f^2+g^2=0, \label{dr4}
\eeq
where 
\beq 
f^2={m^2\over 2}Q\gamma{\eta\over\kappa}={m^2\over 2}{l^2N^2\over n^2\oma^2}{\eta\over\kappa}, \qquad g^2=2m^2h^2.
\eeq
The real part of (\ref{dr3}) yields:
\beq
-2m^2\alpha^3-4m^2\alpha^2+\alpha[2m^2h(h+2k)-(p+1)+{m^2\over 2}(Q+1)]+4mhk=0. \label{dr3r}
\eeq
It is not immediately clear from this equation how the limit $h/k\ll 1$ is to be taken, but multiplying (\ref{dr3i1}) by $\alpha$, subtracting this from (\ref{dr3r}), and multiplying the result by $h/k$ yields:
\beq
\alpha[{m^2\over 2}Q{h\over k}(1-{h\over k})+4m^2h^2]+4mh^2=0. \qquad (\eta/\kappa\ll 1) 
\eeq
Since $Q{h/k}\sim O(1)$ the limit $h/k\rightarrow 0$ thus yields
\beq
m\alpha=-2g^2/(f^2+g^2).
\eeq
This implies that 
\beq 2-2(m\alpha+1)^2>0.\eeq
Hence all terms in (\ref{dr4}) are positive definite except the term involving $p$. A necessary condition for instability is thus that $a\equiv p+1-m^2/2>0$. This is most easily satisfied for $m=1$ (recall that the case $m=0$ has to be treated separately). For azimuthal fields generated by winding up near the pole, $p=1$, hence $a=3/2$ and the condition on $p$ is satisfied for $m=1$. The other {\em necessary} conditions are:
\beq
2-2(m\alpha+1)^2<a\quad{\rm and}\quad f^2<a \quad{\rm and}\quad g^2<a.
\eeq
{\em Sufficient}, on the other hand is the set of conditions
\beq
2-2(m\alpha+1)^2<a/3\quad{\rm and}\quad f^2<a/3 \quad{\rm and}\quad g^2<a/3.
\eeq
The necessary are close enough to the sufficient conditions. The gap between the two can be narrowed with a more detailed analysis but the present approximation is sufficient for application to the dynamo estimates in S02. One then finds that, up to a factor of order unity, a necessary and sufficient condition for instability is (S99 eq. A29:)
\beq 
\oma^4\ga {\vert m\vert^3\over a^{3/2}}l^2N^2{\eta\over\kappa}\eta\Om, \qquad (\kappa\gg\eta,~\Om\gg\oma,\quad p>-1/2,\quad m>0).
\eeq

One may wonder what has happened to the two roots that have disappeared from the original 5th order equation, and whether those could become unstable as well. In the limit $\Om\gg\oma$, these roots turn out to scale as $\omega\sim\Om$. Under this assumption, and expanding in $\epsilon$ as above, the dispersion reduces to
\beq
\omega^2-{l^2\over n^2}N^2-4\Om^2=0.
\eeq
The magnetic field has thus disappeared from the equation, showing that the two remaining roots purely hydrodynamic. They describe inertial-gravity waves, and are unstable only for $N^2<0$, i.e. in convective zones.

\subsection{The 'slowly rotating' case}
\label{lowom}
One may wonder if the assumption $\Om\gg\oma$ is always justified, and how the instability would be different in the opposite case. This question is also motivated by the fact that the dynamo argument in S02 does not lead to a properly defined amplitude for the magnetic field generated in this case. 

It is now more convenient to write the dispersion relation in terms of a dimensionless growth rate $\sigma$ (which, for the moment, is still a complex number):
\beq
\omega=-i\sigma\oma.
\eeq 
Setting 
\beq 
\delta=\Om/\oma, \qquad H\equiv\eta s^2/\oma, \qquad K\equiv\kappa s^2/\gamma\oma,
\eeq
the dispersion relation (\ref{dr1}) can be written as
\begin{eqnarray}
&(\sigma+im\delta)-{m^2\over 2}[\sigma+{1\over\sigma-H}+{Q\over\sigma-K}] [\sigma(\sigma-H)+1] \nonumber\\
&+(2im\delta-{p+1\over\sigma-H}) [im\delta(\sigma-H)-1]=0. \label{drs1}
\end{eqnarray}
To lowest order in $\delta$, the coefficients of this equation are real. The condition for exponential instability is found from the sign of the constant term after multiplying by $(\sigma-H)(\sigma-K)$, and is:
\beq p>-1+{m^2\over 2}(1+Q{\eta\over\kappa}).\eeq
Since $Q$ can be made arbitrarily small by choice of the wavenumbers, there are unstable modes when $p>-1+{m^2/2}$. This condition is as before in the rapidly rotating case: the (logarithmic) gradient in field strength required for instability is the same. At this lowest order in $\Omega/\oma$, there is no threshold for instability for an azimthal magnetic field in a nonrotating star; the only requirement is that the gradient index $p$ be large enough. To find the minimum field strength for instability, the next order in the small quantity $\delta$ has to be considered. With a bit of analysis one finds
\beq
p>-1+{m^2\over 2}+8m\delta,
\eeq
i.e. a small correction for $\delta\ll 1$. More significantly one finds that there is now a minimum field strength for instability:
\beq 
{\oma\over\Om} \ga ({N\over\Om})^{3/4} ({\eta\over\kappa})^{1/4} ({\eta\over r^2N})^{1/4}.\qquad (\kappa\gg\eta) \label{minB}
\eeq
This is the same condition as for the opposite limiting case $\oma/\Om\ll 1$ (see eq.\ (A29) in S99, noting that the equivalent eq.\ (49) in S99 contains a typo, as explained in a footnote in S02). Thus (\ref{minB}) happens to apply irrespective of $\Om$.

\subsection{Relevance of the `slowly rotating' case}

Next, we can ask when the case $\Om\ll\oma$ is likely to apply. For this, we need an estimate of the amplitude of the magnetic field. If it is generated by the dynamo process envisaged in S02, and if the estimate for its amplitude given there is valid, the field strength in the rapidly rotating case would be given by
\beq
{\oma\over\Om}\approx q^{1/2}\left({\Om\over N}\right)^{1/8} \left({\kappa\over r^2 N}\right)^{1/8}, \label{case1}
\eeq
(S02 eqs. 17, 19), where $q$ is the dimensionless rotation gradient $q={\rm d}\ln\Om/{\rm d}\ln r$. This estimate applies if the buoyancy frequency $N$ is dominantly due to the thermal stratification (in a sense that can be defined more precisely). If, instead, the buoyancy due to a gradient of composition is more important (`case 0' in S02), the field strength is given by
\beq
{\oma\over\Om}\approx q{\Om\over N},\label{case0}
\eeq
where $N=N_\mu$ is now the buoyancy frequency due to the stratification of composition.
For the slowly rotating case $\oma/\Om>1$ to apply, the field strength generated by the dynamo has to be high enough. From eqs. (\ref{case1}), (\ref{case0}) it follows that this is possible only for steep rotation gradients $q\gg 1$. Such steep gradients can occur in two cases. 

In normal (slow) stellar evolution, the high effectiveness of the magnetic torques keeps the level of differential rotation low wherever the composition is uniform. The gradients in rotation rate then concentrate in the $\mu$-gradients (e.g.\  Heger et al. 2005). Thus, the most likely place where the slowly rotating case might apply is when the composition gradient determines the buoyancy. Eq. (\ref{case0}) then shows that for $\oma$ to be larger than $\Om$, one must have $q>N/\Om$, or
\beq r{\rm d}\Om/{\rm d}r>N. \eeq
Up to a numerical factor of order unity, this is the Richardson condition for (adiabatic) hydrodynamic instability in a stable stratification. The `slowly rotating' case $\Om\ll\oma$ is thus of limited relevance, since where it applies, ordinary hydrodynamic instabilities will already be effective.

In the above I have assumed that the evolution of the rotation profile in the star is slow enough that the dynamo process is in a quasistationary state, so that winding-up of the field by differential rotation is balanced by its dacay due to magnetic instabilities. If, instead, the rotation profile evolves faster than the time scale on which the dynamo attains its equilibrium amplitude, the field can increase to larger values, solely by the rapid winding-up of field lines. This might occur in pre-supernova cores.

\section{Instabilities in a $\mu$-gradient}
\label{compo}
If the buoyancy is due to a composition gradient, it can not be reduced by thermal diffusion. The only relevant diffusion affecting the buoyancy is then the diffusivity, $\lambda$, of the ions responsible for the stability of the stratification. This diffusivity is generally several orders of magnitude smaller than the magnetic diffusivity (which is mediated by the electrons). This case, $\lambda\ll\eta$, is thus equivalent to setting $\kappa/\gamma=\lambda\ll\eta$ in the (\ref{dr1}), while interpreting $N$ as the buoyancy frequency $N_\mu=(g\partial_r\ln\mu)^{1/2}$ due to the $\mu$-gradient alone. 

In this case, the agent driving the instability (the magnetic configuration) has the higher diffusivity, the composition gradient causing the stabilizing stratification the lower diffusivity of the two. As in the case of other double-diffusive instabilities (e.g.\ Turner 1980), we should expect the existence of an intrinsically oscillatory instability (i.e. even when rotation is absent). An example is the case of hot salty water under cooler fresh water, such that the overall density stratification is stable. Internal gravity waves in this stratification can grow on wavelengths such that the thermal diffusion time is comparable with the wave frequency. The astrophysical equivalent is `semiconvection', where the corresponding linearly unstable modes are called Kato oscillations (Kato 1966). Such modes should also be expected in the present case. 

The analysis is a bit more complicated than the case $\kappa\gg\eta$. To illustrate the nature and stability conditions for such modes, consider the simplified case $\Om=0$. As in the case of Kato oscillations, we expect the unstable oscillations to have frequencies near the buoyancy frequency $F$ of a  wave with vertical and horizontal wavenumber components $n,l$. Introduce the scalings and notation
\beq 
F={l\over n}N,\qquad {\oma\over F}=f,\qquad {\om\over F}=\nu,\qquad h={\eta s^2\over F},\qquad {\lambda s^2\over F}=k
\eeq
(noting that the symbols $h,k$ have been recycled from previous usage above). The dispersion relation (\ref{dr1}) then becomes
\beq
f^2\nu(p-1)+{m^2\over 2}[\nu-{f^2\over\nu+ih}-{1\over\nu+ik}][\nu(\nu+ih)-f^2]-{f^4\over\nu+ih}(p+1)=0.
\eeq
The conditions for marginal overstability are found by assuming $\nu$ real, and taking the Re,Im parts of this equation. The real part has a common factor $\nu$ and after division by this factor reads
\begin{eqnarray}
f^2(p-1)(\nu^2-hk)+{m^2\over 2}[\nu^2-hk-f^2-1](\nu^2-f^2)&\cr
-{m^2\over 2}[\nu^2(h+k)-f^2k-h]h-f^4(p+1)&=0.\label{re}
\end{eqnarray}
This is a quadratic in $\nu^2$. The Im part yields
\begin{eqnarray}
f^2\nu^2(p-1)(1+{k\over h})+{m^2\over 2}[\nu^2(1+{k\over h})-f^2{k\over h}-1](\nu^2-f^2)  &\cr
-{m^2\over 2}[\nu^2-hk-f^2-1]\nu^2-f^4(p+1){k\over h}&=0,\label{im}
\end{eqnarray}
also a quadratic in $\nu^2$.

So far, no assumptions about large and small have been made. The quantity $f$ measures the strength of the magnetic field relative to the stratification. For $f\ga1$, the field is strong enough to overcome the stable stratification irrespective of diffusive effects. Since we are interested in just these effects, we may assume the limiting case $f\ll 1$. By assumption, we also have $\lambda/\eta= k/h\ll1$. Not making further assumptions about the relative smallness of $f$ and  $k/h$, and choosing wavenumbers such that $h^2$ is also small, we set
\beq f^2=\epsilon, \qquad k/h\sim O(\epsilon),\qquad h^2\sim O(\epsilon).\eeq
To order zero in $\epsilon$, both (\ref{re}) and (\ref{im}) then yield $\nu^2=1$, so 
\beq \nu^2=1+O(\epsilon).\eeq
The mode frequency is thus known now, and confirms expectation that it is the frequency of a buoyancy wave. We are thus looking for the conditions under which a buoyancy wave can become overstable. To find the stability condition, the next order in $\epsilon$ has to be considered. It is found by eliminating the term in $\nu^4$ between (\ref{re}) and (\ref{im}). Expanding the result to lowest order in $\epsilon$ then yields
\beq p-1-{m^2\over 2}(1+{k\over hf^2})=0,\label{oco}\eeq
as the condition that holds at marginal oscillatory instability. Evidently, a necessary condition for instability is thus
\beq p>1+{m^2\over 2}.\eeq
Setting $a\equiv p-1-m^2/2$, (\ref{oco}) becomes, in dimensional quantities:
\beq 
\oma^2={l^2\over n^2}N^2{m^2\over 2a}{\lambda\over\eta}\qquad ({\lambda\over\eta} \ll 1,\quad a>0).\label{mnB}
\eeq
Though we formally do not know from this analysis on which side of equality the instability lives, it clear that  (\ref{mnB}) actually defines a minimum field strength for instability, since the field is the ingredient driving the instability.

Condition (\ref{mnB}) is similar to the opposite case $\eta/\kappa\ll 1$, but with $\eta$ and $\kappa$ reversed. This is characteristic of double diffusive instabilities in general. Note, however, that the requirement on the field gradient index $p$ is different from the case $\eta/\kappa\ll 1$, where we had 
$p>-1+{m^2/ 2}$ instead. The oscillatory instability in a $\mu$-gradient thus requires a significantly steeper field gradient. In particular a field would up by differential rotation near the pole, $p=1$, is stable. Also note that the terms $h^2/f^2$ have canceled in the result (\ref{mnB}). This has the effect that there is actually no minimum wavenumber for instability, and hence no minimum field strength either, since  $l/n$ can be taken arbitrarily small in (\ref{mnB}). This differs from the case $\eta/\kappa\ll1$, where at the same order there is a minimum field strength (cf eq. \ref{minB}). It must be noted, however, that the absence of a minimum field strength is probably an artefact of the inviscid approximation used here. Viscosity is likely to define a minimum field strength. In view of the uncertain relevance of this instability, and the likelihood that linear stability gives a misleading impression anyway in this case (see section \ref{layer}), this is not explored further here.

\subsection{layering}
\label{layer}
If a $\mu$-gradient is initially strong enough to prevent instability, evolution of the star will tend to increase the gradient in rotation rate across it until instability happens, either by the process described above, or the simpler form of instability on which the dynamo estimates of S02 are based. Once this happens, however, it is quite possible that the subsequent nonlinear development will lead to {\em layer formation}: the gradient in  $\mu$ breaks up into a series of steps, each with (nearly) uniform composition and rotation rate, separated by steep gradients in $\mu$ and $\Om$. If this happens, the angular momentum transport can potentially increase strongly. 

To see this, imagine replacing the smooth gradient in $\Om$ and $\mu$ by a series of steps of thickness $d$, with boundary layers of some width between the steps. Inside a step, the stabilization by a $\mu$-gradient has disappeared, so that instability and dynamo action can take place as if there were no gradient in composition at all. If the instability time scale is $\sigma^{-1}=\Om/\oma^2$ (see section \ref{nog} above), there will be magnetic boundary layers of width $\delta=(\eta/\sigma)^{1/2}$ bounding each step. Across these very steep boundary layers, the exchange of magnetic flux is entirely by magnetic diffusion rather than fluid motion. 

To set up such a layered structure, energy has to be spent to overcome the stable composition gradient, but once it is in place, it can persist. The energy required to overturn the stratification into layers of thickness $d$ is, per unit mass, $e\approx d^2N_\mu^2$ (equal to the kinetic energy of a vertical buoyancy oscillation of amplitude $d$). If the composition gradient extends over a radial distance $D$, this energy is smaller by a factor $(d/D)^2$ than the energy per unit mass needed to overturn the stratification globally. This small amount of energy might be supplied by, for example, gravity waves from a nearby convective zone, or else by magnetic instability when the gradient in rotation rate has become strong enough in the course of evolution of the star. Once layering sets in, the stabilizing effect of the composition gradient is reduced. This is quite analogous to the case of double diffusive instabilities in geophysics and in the laboratory (c.f.\ Turner 1980, 1985). 

An astrophysical equivalent of this nonlinear, {\em subcritical} behavior of instabilities in a composition gradient is semiconvection. Once a layer within a composition gradient has been overturned, be it by a numerical artefact or by an actual physical process, the mixed state of the layer will be maintained by convection. This is often observed in semiconvective regions of stellar models, even if layer formation (Spruit 1992) is not taken into account explicitly. 

The same may be taking place in stellar evolution calculations which include magnetic torques according to S02. In the results of Heger et al.\ (2005), for example, there is evidence of spiky structure in the $\mu$-gradients on length scales comparable to the numerical resolution.  To assess its consequences quantitatively, it would be important to find ways of taking the layering process into account in a more controlled way.

\end{document}